\documentclass[twoside,twocolumn,9pt]{article}
\usepackage{extsizes}
\usepackage[super,sort&compress,comma]{natbib} 
\usepackage[version=3]{mhchem}
\usepackage[left=1.5cm, right=1.5cm, top=1.785cm, bottom=2.0cm]{geometry}
\usepackage{balance}
\usepackage{mathptmx}
\usepackage{sectsty}
\usepackage{graphicx} 
\usepackage{lastpage}
\usepackage[format=plain,justification=justified,singlelinecheck=false,font={stretch=1.125,small,sf},labelfont=bf,labelsep=space]{caption}
\usepackage{float}
\usepackage{fancyhdr}
\usepackage{fnpos}
\usepackage[english]{babel}
\addto{\captionsenglish}{%
  \renewcommand{\refname}{Notes and references}
}
\usepackage{array}
\usepackage{siunitx}
\usepackage{adjustbox}
\usepackage{multirow}
\usepackage{droidsans}
\usepackage{charter}
\usepackage[T1]{fontenc}
\usepackage{setspace}
\usepackage[compact]{titlesec}
\usepackage{hyperref}
\usepackage{verbatim}
\usepackage{xcolor}
\usepackage{textcomp}    
\usepackage{amsmath}     
\usepackage{gensymb}     
\usepackage{stfloats}

\newcommand{\credit}[2]{\textbf{#1:} #2\\}

\usepackage{epstopdf}

\definecolor{cream}{RGB}{222,217,201}

\begin{document}

\pagestyle{fancy}
\thispagestyle{plain}
\fancypagestyle{plain}{
\renewcommand{\headrulewidth}{0pt}
}

\makeFNbottom
\makeatletter
\renewcommand\LARGE{\@setfontsize\LARGE{15pt}{17}}
\renewcommand\Large{\@setfontsize\Large{12pt}{14}}
\renewcommand\large{\@setfontsize\large{10pt}{12}}
\renewcommand\footnotesize{\@setfontsize\footnotesize{7pt}{10}}
\makeatother

\renewcommand{\thefootnote}{\fnsymbol{footnote}}
\renewcommand\footnoterule{\vspace*{1pt}%
\color{cream}\hrule width 3.5in height 0.4pt \color{black}\vspace*{5pt}} 
\setcounter{secnumdepth}{5}

\makeatletter 
\renewcommand\@biblabel[1]{#1}            
\renewcommand\@makefntext[1]%
{\noindent\makebox[0pt][r]{\@thefnmark\,}#1}
\makeatother 
\renewcommand{\figurename}{\small{Fig.}~}
\sectionfont{\sffamily\Large}
\subsectionfont{\normalsize}
\subsubsectionfont{\bf}
\setstretch{1.125} 
\setlength{\skip\footins}{0.8cm}
\setlength{\footnotesep}{0.25cm}
\setlength{\jot}{10pt}
\titlespacing*{\section}{0pt}{4pt}{4pt}
\titlespacing*{\subsection}{0pt}{15pt}{1pt}

\fancyfoot{}
\fancyfoot[RO]{\footnotesize{\sffamily{~\textbar\hspace{2pt}\thepage}}}
\fancyfoot[LE]{\footnotesize{\sffamily{\thepage~\textbar}}}
\fancyhead{}
\renewcommand{\headrulewidth}{0pt} 
\renewcommand{\footrulewidth}{0pt}
\setlength{\arrayrulewidth}{1pt}
\setlength{\columnsep}{6.5mm}
\setlength\bibsep{1pt}

\makeatletter 
\newlength{\figrulesep} 
\setlength{\figrulesep}{0.5\textfloatsep} 

\newcommand{\topfigrule}{\vspace*{-1pt}%
\noindent{\color{cream}\rule[-\figrulesep]{\columnwidth}{1.5pt}} }

\newcommand{\botfigrule}{\vspace*{-2pt}%
\noindent{\color{cream}\rule[\figrulesep]{\columnwidth}{1.5pt}} }

\newcommand{\dblfigrule}{\vspace*{-1pt}%
\noindent{\color{cream}\rule[-\figrulesep]{\textwidth}{1.5pt}} }

\makeatother

\twocolumn[

\sffamily
\begin{tabular}{m{13.5cm} p{4.5cm} }

\noindent\LARGE{\textbf{MultiTaskDeltaNet: Change Detection-based Image Segmentation for \textit{operando} ETEM with Application to Carbon Gasification Kinetics}}  & \\

\vspace{0.3cm} & \vspace{0.3cm} \\

\noindent\large{Yushuo Niu,\textit{$^{a}$} Tianyu Li,\textit{$^{b}$} Yuanyuan Zhu,\textit{$^{b}$} and Qian Yang\textit{$^{*a}$}} & \\

\vspace{0.3cm} & \vspace{0.3cm} \\
\noindent\normalsize 
Transforming in-situ transmission electron microscopy (TEM) imaging into a tool for spatially-resolved \textit{operando} characterization of solid-state reactions requires automated, high-precision semantic segmentation of dynamically evolving features. However, traditional deep learning methods for semantic segmentation often encounter limitations due to the scarcity of labeled data, visually ambiguous features of interest, and small-object scenarios. To tackle these challenges, we introduce MultiTaskDeltaNet (MTDN), a novel deep learning architecture that creatively reconceptualizes the segmentation task as a change detection problem. By implementing a unique Siamese network with a U-Net backbone and using paired images to capture feature changes, MTDN effectively utilizes minimal data to produce high-quality segmentations. Furthermore, MTDN utilizes a multi-task learning strategy to leverage correlations between physical features of interest. In an evaluation using data from in-situ environmental TEM (ETEM) videos of filamentous carbon gasification, MTDN demonstrated a significant advantage over conventional segmentation models, particularly in accurately delineating fine structural features. Notably, MTDN achieved a $10.22\%$ performance improvement over conventional segmentation models in predicting small and visually ambiguous physical features. This work bridges several key gaps between deep learning and practical TEM image analysis, advancing automated characterization of nanomaterials in complex experimental settings.

\end{tabular}

 
 \vspace{0.6cm}

  ]

\renewcommand*\rmdefault{bch}\normalfont\upshape
\rmfamily
\section*{}
\vspace{-1cm}


\footnotetext{\textit{$^{a}$School of Computing, University of Connecticut, Storrs, CT, USA; E-mail: qyang@uconn.edu}}
\footnotetext{\textit{$^{b}$Department of Materials Science and Engineering, University of Connecticut, Storrs, CT, USA}}



\section{Introduction}
\textit{Operando} transmission electron microscopy (TEM) has recently emerged as a transformative technique in materials characterization by enabling in-depth investigations into the kinetics and mechanisms of structural, morphological, and phase transformations~\cite{Chee2023,Yang2023,Zheng2021}. Building on in-situ TEM, operando TEM simultaneously measures materials functionality (e.g., phase transformation reactivity) alongside in-situ imaging, thereby facilitating quantitative correlations between microstructural evolution and reaction kinetics. Specifically, for gas-solid reactions studied using \textit{operando} environmental TEM (ETEM)~\cite{jinschek2024quantitative}, in-situ reactivity measurements are often performed by monitoring reactant and product gases or solid phases using auxiliary mass spectrometry (MS)~\cite{Miller2014,Vendelbo2014}, electron energy loss spectroscopy (EELS)~\cite{Chenna2012,Jeangros2014}, or selected area electron diffraction (SAED)~\cite{Yu2018,sainju2022defecttrack}. One of the grand challenges in \textit{operando} ETEM studies is the difficulty of precisely correlating spatiotemporal structural \textbf{\underline{changes}} with their corresponding reaction kinetics~\cite{Chee2023}. While the current spatial and temporal resolutions of in-situ imaging employed in conventional TEM are sufficient to capture the microstructural evolution at nanoscale for a broad range of solid-state reactions such as nanomaterials nucleation, growth, oxidation and reduction, \textit{operando} ETEM employing conventional spectroscopic or diffraction techniques provides only averaged in-situ reactivity measurement. Consequently, these techniques lack the spatial resolution required to reliably connect reaction kinetics with microstructural evolution of individual nanostructures, which often exhibit size or structural heterogeneities.

Semantic segmentation—a pixel-level classification task in computer vision~\cite{horwath2020understanding}—is well-suited for quantifying temporal changes in feature size from in-situ ETEM videos. In our previous studies of nanostructure phase transformations, manual segmentation allowed us to obtain spatially-resolved reaction kinetics, providing unprecedented insights into size-dependent oxidation of Ni nanoparticles~\cite{sainju2022defecttrack}, quantitative comparison of competing reaction pathways during filamentous carbon gasification~\cite{Nielsen2023,Nielsen2025}, and unexpected irradiation-decelerated tungsten nanofuzz oxidation that challenges conventional understanding~\cite{Sainju2024}. However, manual segmentation is labor-intensive and limits scalability, underscoring the need for automated approaches to enhance statistical power and standardization.

Recent advances in deep learning, particularly convolutional neural networks (CNNs) including U-Net and transformer-based architectures such as Vision Transformer (ViT), have revolutionized segmentation tasks in many fields\cite{horwath2020understanding,long2015fully,milletari2016v,ronneberger2015u,dosovitskiy2021an,Yu2015MultiScaleCA,chen2018encoder,he2016deep,zhang2018road,huang2017densely,li2018h,roberts2019deep,howard2017mobilenets,zunair2021sharp,woo2018cbam,oktay2018attention,chen2021transunet,cao2022swin}. However, segmentation of microscopy videos remains challenging due to limited annotated datasets, complex image features which differ significantly from natural images in texture and scale, and the presence of small and/or ambiguous objects~\cite{ziatdinov2017deep}. Foundation models such as the Segment Anything Model (SAM)\cite{kirillov2023segment} offer zero-shot segmentation but struggle to generalize to scientific domains without extensive domain-specific data for fine-tuning or high-quality prompts\cite{mazurowski2023segment,ma2024segment,li2024polyp,wu2025medical}. Self-supervised learning methods like SimCLR and Barlow Twins can help address labeled data scarcity\cite{chen2020simple,lu2022semi,zbontar2021barlow} but themselves require large amounts of unlabeled data to be effective, especially for segmentation of complex images~\cite{konstantakos_self-supervised_2025,xie_data_2022,el-nouby_are_nodate}.

To develop automated and reliable segmentation models for microscopy videos, we adopt the \textit{operando} ETEM gasification of filamentous carbon as a model system to identify the specific challenges and current domain needs. Understanding filamentous carbon gasification is critical for gaining fundamental insights into catalyst regeneration mechanisms, enabling the development of more effective strategies to restore catalyst activity from coking - the leading cause of deactivation in thermal heterogeneous catalysis~\cite{martin2022unifying}. As shown in Fig. 1a, microelectromechanical system (MEMS)-based ETEM experiments were conducted to emulate high-temperature carbon gasification under industrially relevant air-like conditions. An in-situ ETEM video captured the dynamic behavior and gradual removal of over 100 filamentous carbon, revealing complex gasification phenomena involving three competing reaction pathways~\cite{Nielsen2025}. For example, the classic catalytic gasification pathway is presented in Fig.~\ref{fig:experiment}a. Although combining built-in mass spectrometry (MS) with in-situ ETEM observations provides viable \textit{operando} characterization, MS measures the total gas products at the ETEM cell outlet, yielding only averaged gasification kinetics across mixed filamentous carbon sizes and reaction pathways. Therefore, a spatially-resolved method is needed to measure individual filament-level (i.e. filament-specific) gasification kinetics and thus deconvolute the mixed contributions, enabling quantitative comparison among the three gasification pathways.

Three main challenges hinder automated segmentation in this domain. Firstly, there is currently no open-source benchmark database of professionally annotated in-situ (E)TEM videos. Often, only a limited set of ground-truth labeling data specific to particular nanostructures and reactions is available for machine learning model training. This creates a ``small data" problem for training deep learning based models, which typically need large, pixel-level annotated datasets that are labor-intensive and require domain expertise to obtain\cite{wang2021pairwise}.

Secondly, to facilitate spatially-resolved reaction kinetics extraction from in-situ ETEM videos, segmentation focuses on `reactivity descriptors' of nanostructures rather than apparent image features. In this case (Fig.~\ref{fig:experiment}b), following the convention in dedicated ex-situ gasification kinetic tests~\cite{Alenazey2009}, filamentous carbon volume should be quantified as a function of gasification reaction time. This requires segmentation of two `reactivity descriptors': A$_1$ (the entire carbon projection area) and A$_2$ (the hollow core area) of the multiwall carbon nanotube (MWCNT)-like filamentous carbon observed in this spent Ni catalyst~\cite{Nielsen2023}, which are then used to quantify volume changes using an area-to-volume conversion (Fig.~\ref{fig:experiment}b). The visual similarity of $A_2$ to the background is challenging for general-purpose segmentation models.

Thirdly, segmentation tasks in this domain unavoidably involve ``small objects"~\cite{Tong2020}—whether emerging reaction products that start small at early reaction stages (e.g., MWCNT growth) or solid reactants such as filamentous carbon, which become increasingly small towards the end of the reaction. This is particularly challenging for our `reactivity descriptor’ A$_2$, as it begins as a small object.

Finally, additional complications, including overlapping nanostructures and feature blur due to rapid motion, further complicate segmentation. While physics-based machine learning models have been proposed as an attractive approach, they hinge on validated, known kinetic models that are frequently unavailable or untested at the nanoscale~\cite{sainju2022defecttrack}.

To address these challenges in quantifying object evolution in microscopy video data, especially object size, we introduce MultiTaskDeltaNet (MTDN), a deep learning model tailored for filamentous carbon segmentation in ETEM videos. The key innovation of MTDN is to reframe the segmentation problem as a change detection task, by leveraging a Siamese architecture with pairwise data inputs to augment limited training data and improve generalization. A lightweight backbone, combined with pre-training and fine-tuning strategies, ensures efficiency while maintaining high performance. The model also employs a multi-task learning framework to simultaneously segment both reactivity descriptors A$_1$ and A$_2$, using their spatial and structural correlation to boost accuracy, especially for the more challenging A$_2$ region. This approach is the first, to our knowledge, to robustly segment both filament areas in low-resolution ETEM videos, enabling detailed analysis of nanoscale carbon gasification kinetics.

\section{Method}
In the following sections, we will describe how the dataset is processed to enable reframing of segmentation as a change detection task, as well as the corresponding MultiTaskDeltaNet model architecture.

\subsection{Dataset}
\subsubsection{Ground Truth Labeling}
\begin{figure*}[!h]
    \centering 
     \includegraphics[width=0.70\textwidth]{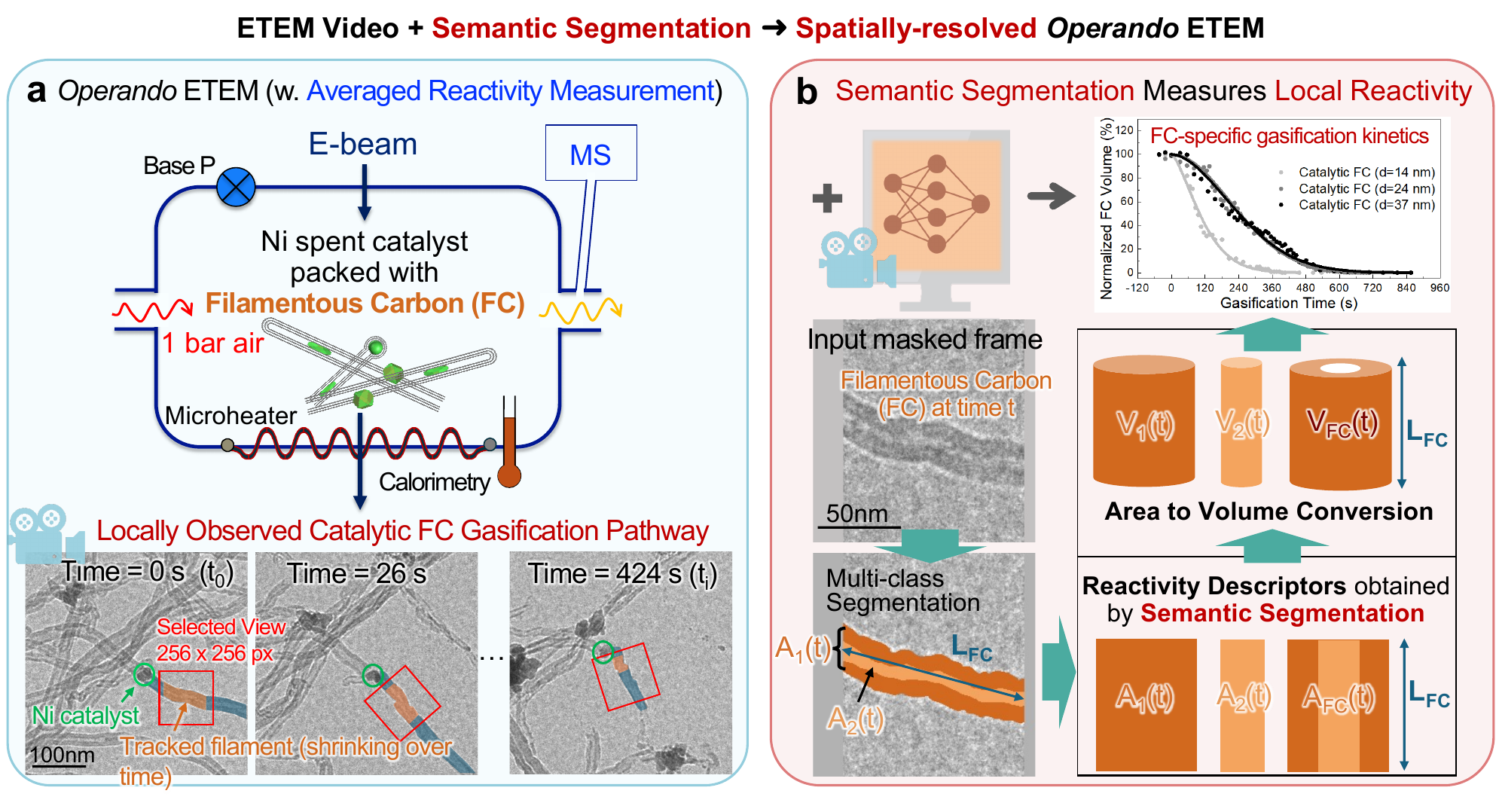}   
    \caption{Schematic overview of the spatially-resolved operando ETEM used to study filamentous carbon gasification. (a) Conventional ETEM setup and an example of the catalytic carbon gasification mode. (b) Semantic segmentation enables spatially-resolved reactivity measurement. Using filamentous carbon gasification as a model system, we segment two ``reactivity descriptors”, A$_1$ (the entire filament projection area) and A$_2$ (the hollow core area), to quantify changes in carbon volume for specific filament size and/or gasification mode.}
    \label{fig:experiment}
\end{figure*}

For this study, we applied the following steps to produce time-dependent filament-specific ground truth labeling. First, an original $4096 \times 4096$ ETEM video was cropped into seven $256 \times 256$ regions (Fig.~\ref{fig:experiment}a), with each region centered on a primary carbon filament for segmentation. The $256 \times 256$ input size is commonly adopted in computer vision benchmarks and compatible with standard deep learning architectures. Next, non-target filament and other objects within the cropped region were masked out as background (shown in grey) to generate the ``masked frames" used as inputs to our model training (Fig.~\ref{fig:dataset}). Then, two researchers with extensive experience in bright-field TEM (BF-TEM) jointly annotated the reactivity descriptors A$_1$ and A$_2$ for each of the seven target filaments. Depending on the filament’s gasification progress, cropped video frames were sampled every 20 to 60 seconds, yielding 14 to 51 frames per filament (Fig.~\ref{fig:dataset}). Using the GNU Image Manipulation Program (GIMP), a total of 231 video frames were annotated by iteratively tracking each filament and cross-examining the ground-truth labels over multiple passes.

\subsubsection{Data Partitioning}
\begin{figure}[!h]
    \centering 
     \includegraphics[width=0.48\textwidth]{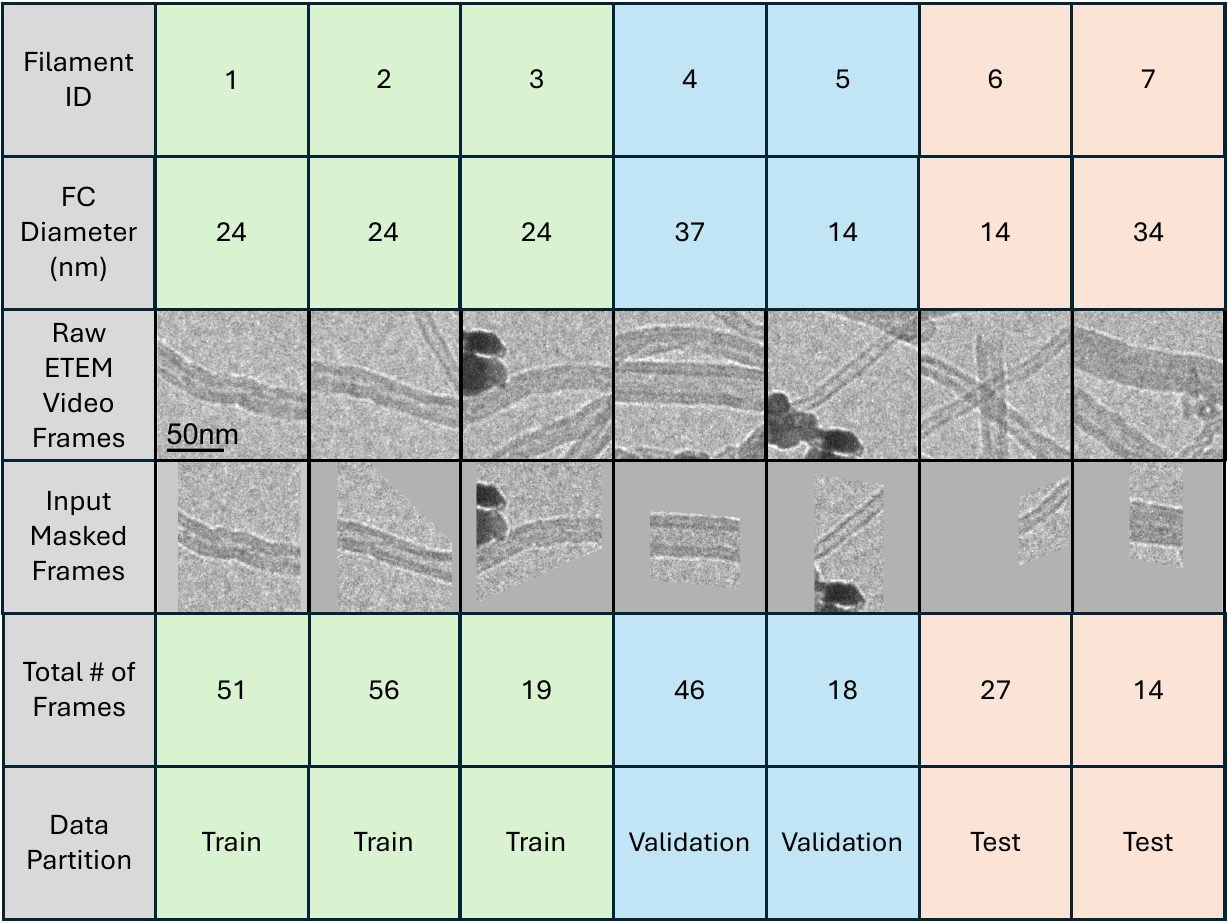}   
    \caption{Summary of annotated ETEM video frames and data partitioning. Carbon filaments of 24 nm diameter (IDs 1-3) were selected for training, representing the most common FC size. Filaments with varying diameters (14–37 nm) were used for validation (IDs 4–5) and testing (IDs 6–7) to evaluate model generalizability across filament sizes. In total, 231 annotated frames were used.}
    \label{fig:dataset}
\end{figure}

As shown in Fig.~\ref{fig:dataset}, the full dataset comprises 231 labeled ETEM frames for seven carbon filaments with diameters ranging from 14 nm to 37 nm. Since the majority of the filamentous carbon in our ETEM gasification study measure around 24 nm in diameter~\cite{Nielsen2025}, we selected filaments 1–3 (each 24 nm) as the training set to represent the most common object size, totaling 126 frames. To ensure spatial separation and prevent data leakage, filaments 4 and 5 (37 nm and 14 nm) were assigned to the validation set (64 frames), and filaments 6 and 7 (14 nm and 34 nm) were reserved for testing (41 frames). This data partitioning allows us to evaluate the model’s ability to generalize across the full range of filament sizes observed.

\subsubsection{Pairwise Change Detection Dataset}
Pairwise image modeling has gained considerable attention in unsupervised video object segmentation (VOS) to capture relationships between frames, often with Siamese networks and attention mechanisms\cite{lu2020zero,zhang2021new,wang2021pairwise,xie2021learning}. It is also a popular approach for change detection tasks\cite{niu2023semi,kim2025learning}. However, these existing methods are not designed for supervised semantic segmentation, particularly when it comes to segmenting visually ambiguous features such as the reactivity descriptor A$_2$ in our problem setting.

\begin{figure}[!h]
    \centering 
     \includegraphics[width=0.4\textwidth]{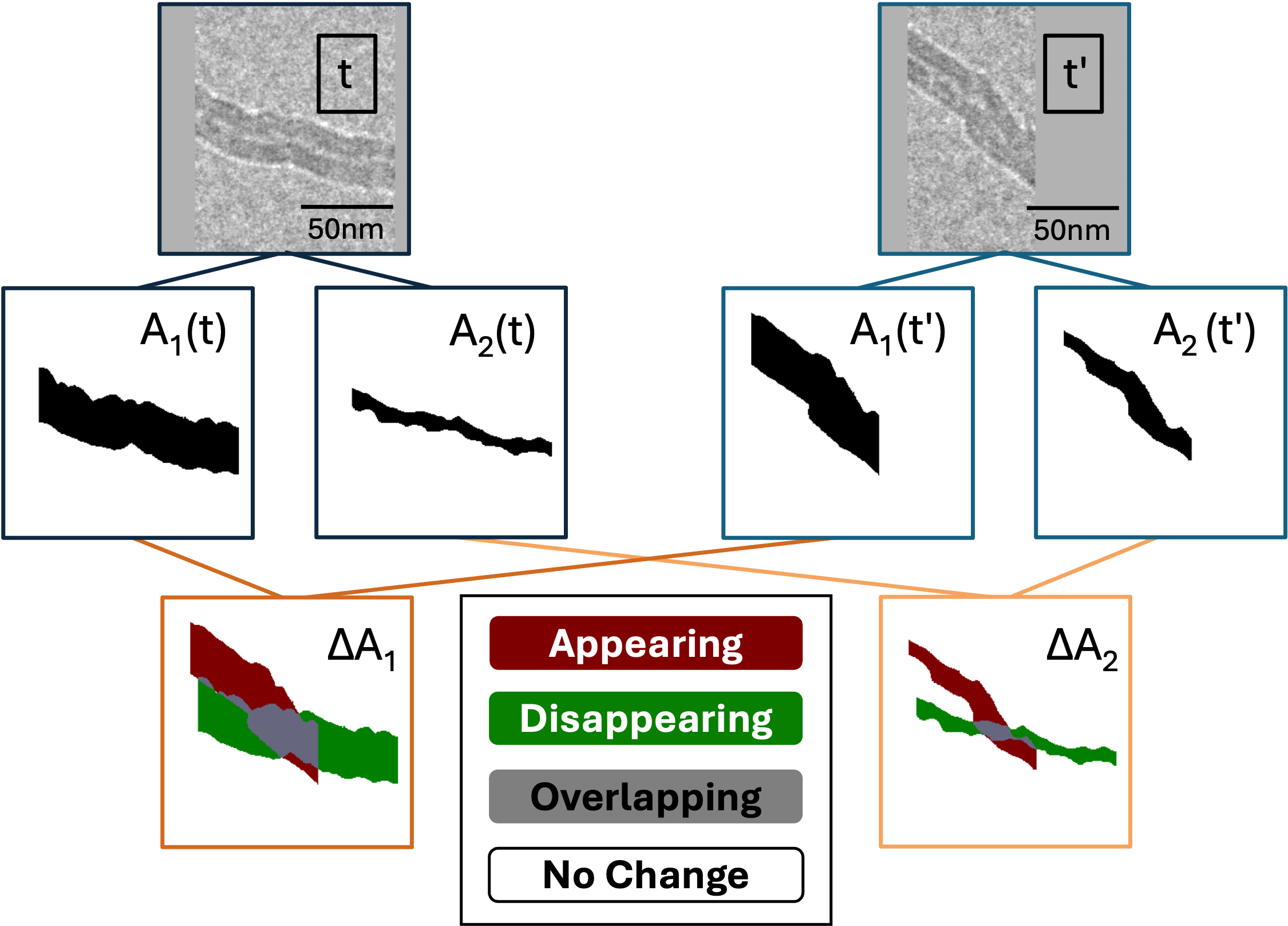}   
    \caption{Schematic of the pairwise data and change detection label generation. The dataset originates from a segmentation task involving a single frame with two segmentation labels: reactivity descriptors $(A_1)$ and $(A_2)$. To adapt this for change detection, frame pairs are taken at different time steps ($t$ and $t{\prime}$), and the corresponding change detection labels ($\Delta A_1$, $\Delta A_2$) are derived. For each pixel in $A_1$ (and similarly for $A_2$), change detection labels are assigned based on comparisons between $t$ and $t{\prime}$: \underline{Appearing}: Pixel is present in $A_1(t{\prime})$ but not in $A_1(t)$,\underline{Disappearing}: Pixel is present in $A_1(t)$ but not in $A_1(t{\prime})$,\underline{Overlapping}: Pixel exists in the same location in both $A_1(t)$ and $A_1(t{\prime})$, \underline{No Change}: None of the above conditions apply. This process results in two change detection labels ($\Delta A_1$ and $\Delta A_2$) per frame pair to be used in the multi-task model.}
    \label{fig:Pairwise_dataset}
\end{figure}

Here, we use labeled ETEM video frames from our dataset to create a change detection dataset, consisting of pairs of frames and the corresponding pixel-wise segmentation label. We consider all pairs of frames of the same filament at different reaction time steps, \( t \) and \( t' \). For each pair (Fig.~\ref{fig:Pairwise_dataset}), the segmentation label categorizes pixels into one of four categories for each of $i\in\{1,2\}$ corresponding to the reactivity descriptors $A_1$ and $A_2$:

- Category ``\textit{appearing}", if a pixel is present in $A_i(t')$ but not in $A_i(t)$.

- Category ``\textit{disappearing}", if a pixel is present in $A_i(t)$ but not in $A_i(t')$.

- Category ``\textit{overlapping}", if both pixels occupy the same location in both frames.

- Category ``\textit{no change}", if none of the above conditions apply. These correspond to background pixels that remain in the background.

As each frame contains two segmentation labels for two different areas, there are two change detection labels for each frame pair, which we refer to as $\Delta A_1$ and $\Delta A_2$. These change detection labels $\Delta A_i$ can be computed directly from the original frame labels $A_i(t)$ and $A_i(t')$, without any further manual labeling.

Converting our original labeled frame dataset to a pairwise change detection dataset naturally expands the effective size of the dataset, as shown in Table~\ref{tab:pairwise}. This is particularly valuable in low-data environments where manual labeling is costly. Additionally, our pairwise method introduces an implicit regularization effect. The same piece of carbon filament captured at different time steps may exhibit slight variations in size, shape, and position. This variability helps prevent the model from overfitting to specific time step conditions and improves its generalization across various sequences of carbon filament frames. Pairing frames from different time steps within the same region instead of across regions is important, however, to ensure that the model is focused on capturing subtle changes to the object of interest rather than variations in filament diameter and other environmental differences.

\begin{table}[!h]
\caption{The training, validation, and test datasets contain 231, 64, and 41 frames, respectively. The frames in each data partition were then used to generate pairs as described in Section 2.1.3, leading to 2,986, 1,188, and 442 paired data in each partition, respectively.}
\begin{adjustbox}{width=0.55\columnwidth,center}
\begin{tabular}{|c| c| c|} 
 \hline
 \textbf{Dataset} &  \textbf{Original} &  \textbf{Pairwise}  \\
 \hline\hline
 Training     & 231    & 2968  \\  
 \hline
  Validation     & 64    & 1188  \\  
 \hline
   Testing    & 41    & 442  \\  
 \hline
\end{tabular}
\end{adjustbox}
\label{tab:pairwise}
\end{table}

\subsubsection{Recovering Segmentation from Change Detection} \label{sec: area interest output}
\begin{figure*}[t]
    \centering 
     \includegraphics[width=0.7\textwidth]{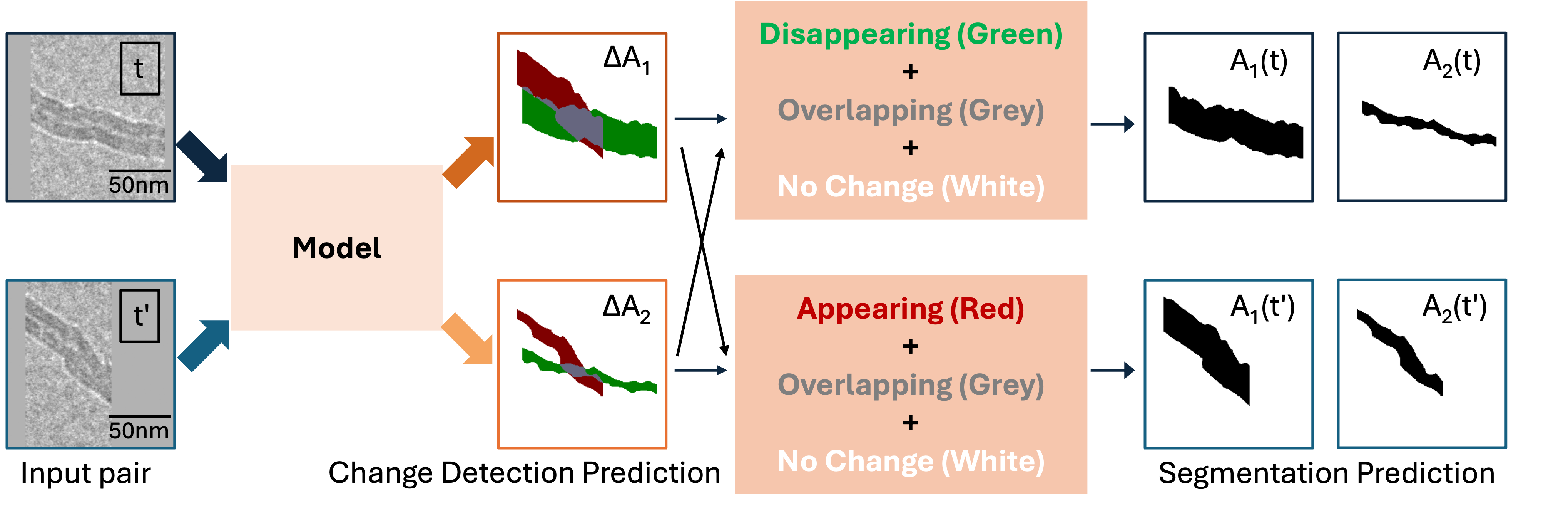}   
    \caption{The segmentation results for each frame are obtained by combining different classes of change detection predictions: For the first frame $img(t)$, segmentation $A_1(t)$ is formed by merging disappearing (green) and overlapping (grey) pixels as the predicted label, with no change (white) pixels as the background in $\Delta A_1$. Segmentation of $A_2(t)$is obtained using the same method on $\Delta A_2$. For the second frame $img(t')$, segmentation of $A_1(t')$ is created by merging appearing (red) and overlapping (grey) pixels as the predicted label, with no change (white) pixels as the background in $\Delta A_2$. Segmentation of $A_2(t')$ is derived using the same approach on $\Delta A_2$. This method reconstructs segmentation results from change detection labels ($\Delta A_1$, $\Delta A_2$) without requiring additional information.}
    \label{fig:CD_to_seg}
\end{figure*}

Our choice of four categories in the change detection labels is also designed to enable the direct recovery of segmentation results from the change detection results $\Delta A_1$ and $\Delta A_2$. Consider the frame pair $img(t)$ and $img(t')$ in Fig.~\ref{fig:CD_to_seg}, where the actual time order of $t$ and $t'$ does not need to be constrained. We can obtain segmentation results for $A_i(t)$ by taking the union of the disappearing and overlapping regions of $\Delta A_i$. Similarly, $A_i(t')$ can be recovered from the union of the appearing and overlapping regions of $\Delta A_i$.

During testing, various methods are possible for transforming a change detection prediction to an image segmentation prediction for a particular time $t$. All methods require inputting two frames from the same filament to the change detection model. The first method is the forward transformation, where each frame is paired with the first frame in time for the same filament. The second method is the backward transformation, which pairs each frame with the last frame in time. The third method is consecutive transformation, where each frame is paired with the next consecutive frame in time. Lastly, the ensemble transformation involves pairing each frame with all other frames corresponding to the same filament, and averaging over the predictions: $T_i=\frac{1}{n} \sum_{n=1}^N T_{in}$, where $T_{in}$ is the transformation based on the pair of frames $img(T_i)$ and $img(T_n)$. Note that self-pairs are included in each of the methods above, allowing the model to be trained with examples of no change. We collectively call these methods \textit{prediction fusion} methods.

\subsection{Siamese Network Architecture }
\begin{figure*}[!t]
    \centering 
     \includegraphics[width=0.75\textwidth]{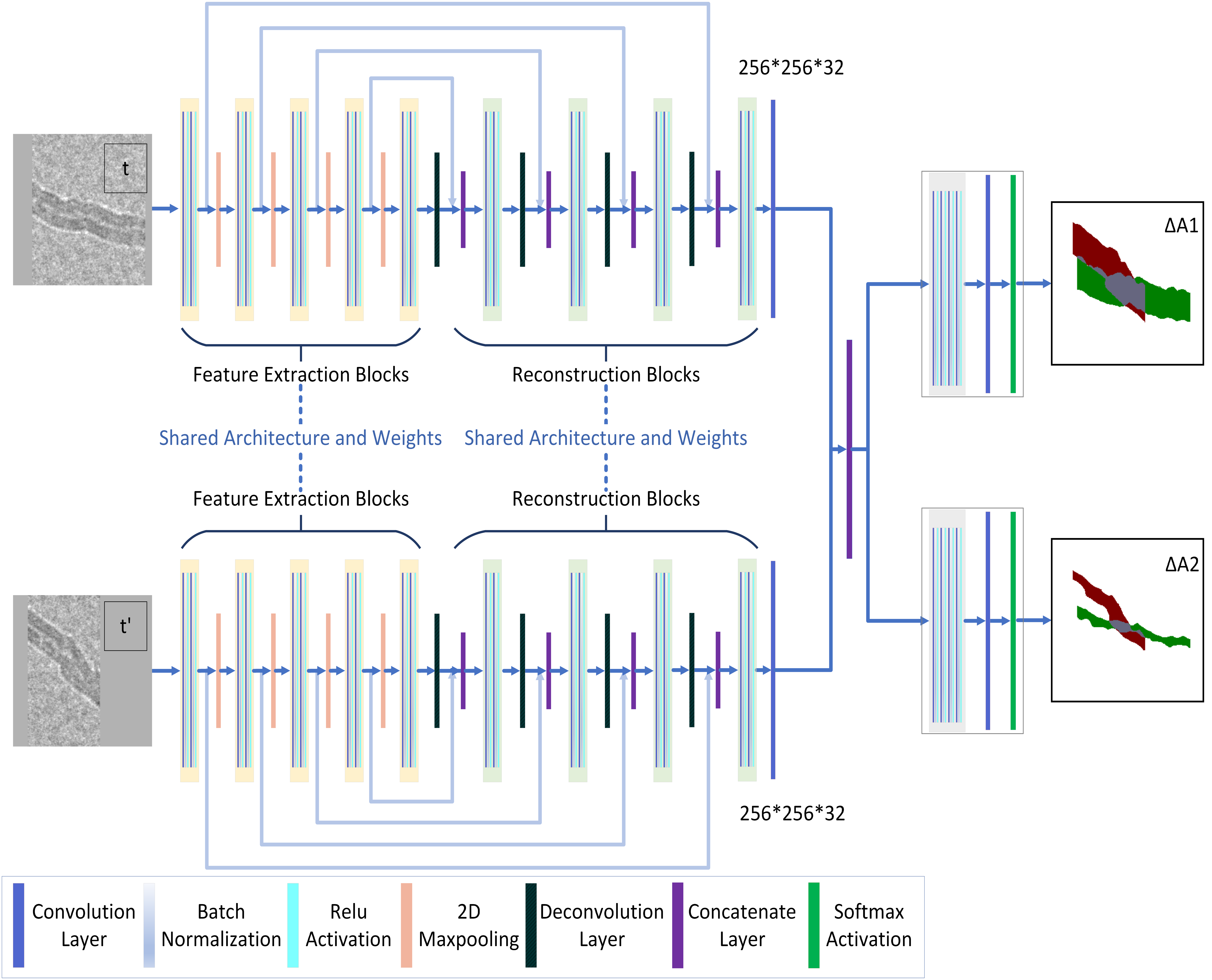}   
    \caption{Architecture of our MultiTaskDeltaNet model. The Siamese branches each incorporate a U-Net backbone, which includes a feature extraction (encoder) section and a reconstruction (decoder) section. Additionally, there are skip connections between corresponding layers to enhance performance. The outputs of the Siamese branches are concatenated before they are input into fully connected layers for each task (A$_1$ change detection and A$_2$ change detection). Different types of layers are color-coded according to the legend.
}
    \label{fig:Model_arch}
\end{figure*}

As shown in Fig.~\ref{fig:Model_arch}, our MultiTaskDeltaNet model consists of two main components: a Siamese architecture based on U-Net branches (although this backbone can be varied), and a set of fully convolutional layers (FCN) for the final change mask generation for each task. The model takes as input a pair of 2D carbon gasification frames from the same region at times $t$ and $t'$. The Siamese architecture features two U-Net branches that share an identical architecture and weights, and first extracts feature maps ($fm (t)$ and $fm (t')$) from each frame. These feature maps are then concatenated, and each FCN then processes the merged features to generate the final change detection output $\Delta A_1$ and $\Delta A_2$, respectively.

We employ pre-training to enhance model performance by training a U-Net model with the same architecture as the backbone of the MTDN branches. This U-Net takes one labeled image frame of filament gasification as its input, and its output is the corresponding segmentation label.

\subsection{Training Objective}
Our model employs focal loss~\cite{lin2017focal} as the loss function to deal with the imbalance in change detection datasets between easy-to-classify background pixels and the smaller number of foreground pixels where changes may occur. The focal loss is a modification of the standard cross-entropy loss designed to address the class imbalance problem.  It allows the model to concentrate more on the hard-to-classify and underrepresented classes while giving less attention to the majority of easily classified classes. The equation for the focal loss is as follows:
\begin{equation}
    \text{FL}(\textbf{p}) = -\sum\limits_{i=1}^{N}\alpha_{i} (1-p_{i})^{\gamma} \log (p_{i})
\end{equation}
where $p_i$ is the predicted probability for the true class $y_i$ for pixel $i$, $\alpha_i$ is a class weighting factor to ensure the loss is not dominated by the majority class, and $\gamma$ is the focusing parameter, which controls the rate at which easy examples are down-weighted.

\subsection{Implementation Details}
Model training was carried out on a Linux system, using Python 3.12, PyTorch 2.0, and CUDA 11.7. The hardware setup included an Nvidia GeForce A5000 GPU. During training, the input image size is set to \(256 \times 256\). For data augmentation, we use vertical and horizontal flips, rotations, image cropping, blurring, and color jittering. The optimizer used is AdamW with \(\beta_1\) at 0.9, \(\beta_2\) at 0.999, and a weight decay of 0.01. The learning rate scheduler decreases the learning rate linearly throughout the training epochs. The default number of epochs is 500, with an early stopping mechanism in place. The learning rate and batch size are determined through hyperparameter tuning using Ray Tune~\cite{liaw2018tune}, with the chosen values being 0.00095 for the learning rate and 16 for the batch size.

\section{Results and discussions}

We compare our results across four models: U-Net\cite{ronneberger2015u}, which is lightweight and the most commonly used segmentation model in scientific applications such as biomedical imagining; our MTDN model trained from scratch, with U-Net as the backbone of its Siamese branches; and two variations of MTDN that initialize training with different pre-trained U-Net weights. All versions of MTDN are designed for change detection, and the simple transformation described in Section~\ref{sec: area interest output} is applied to convert the change detection results into corresponding segmentation results. Consequently, all performance metrics are based on the final segmentation results.

\subsection{Evaluation Metrics}
We use several evaluation metrics to quantitatively assess our model performance. For similarity measures, we include the F1 (Dice) score and Intersection over Union (IoU). These metrics are denoted with an upward arrow $(\uparrow)$ to indicate that higher values represent better performance. The equations for the F1 Score and IoU are as follows:

\begin{equation}
    \text{F1} = \frac{2 \cdot \text{TP} }{2 \cdot \text{TP} + \text{FP} + \text{FN} } = \frac{2 \cdot \lvert \text{GT} \cup \text{Pred} \rvert}{\lvert \text{GT} \rvert + \lvert \text{Pred} \rvert}  
\end{equation}

\begin{equation}
        \text{IoU} = \frac{\text{TP}}{\text{TP} + \text{FP} +\text{FN}} = \frac{\lvert \text{GT} \cup \text{Pred} \rvert}{\lvert \text{GT} \cap \text{Pred} \rvert}
\end{equation}

The final F1 score we report is the macro-averaged F1 score, which computes the F1 score for each class individually and then takes a simple average across classes.

\subsection{Quantitative Results} \label{sec: quantitative Results}
Tables ~\ref{tab: A1 results} and ~\ref{tab: A2 results} compare the performance of the MTDN model and U-Net in predicting the reactivity descriptors $A_1$ and $A_2$ for the test set filaments 6 and 7 (recall Fig~\ref{fig:dataset}). The best performance in each metric is indicated in bold. 

The reactivity descriptor $A_2$ is significantly more challenging than $A_1$ to segment due to its small size and visual similarity to the background. Therefore, improving the prediction performance of $A_2$ is a key focus of our work. MTDN consistently and significantly outperforms U-Net across both metrics (F1 Score and IoU), achieving a $10.22\%$ improvement in F1 Score and a $12.34\%$ improvement in IoU for total $A_2$ prediction. The most notable improvements are seen in IoU, which indicate that MTDN captures object boundaries with greater precision and achieves better overlap with ground truth labels. By contrast, traditional methods such as U-Net performs similarly to our MTDN model on the more straightforward $A_1$ prediction, although MTDN still achieves a slightly higher total test performance.

To place these quantitative results in context, we consider the range of IoU values reported by other works attempting similar segmentation tasks on related types of TEM data. Recently, Yao et al.\cite{yao2020machine} demonstrated that U-Net models trained on simulated liquid-phase TEM data could effectively extract dynamic nanoparticle features from noisy video sequences. They report an optimal IoU of approximately $0.92$ for high signal-to-noise ratio (SNR) images (dose rate = $10\ \mathrm{e^{-} \cdot \mathring{A} ^{-2} \cdot s^{-1}}$) and $0.90$ for low SNR images (dose rate = $1\ \mathrm{e^{-} \cdot \mathring{A} ^{-2} \cdot s^{-1}}$). However, their datasets consisted of high-contrast and relatively simple morphologies, even under low SNR conditions. Similarly, Lu et al.\cite{wang2021pairwise} proposed a semi-supervised segmentation framework for high-resolution TEM images of protein and peptide nanowires. With only eight labeled images per class, they achieved median Dice scores above $0.70$ and IoU values ranging from $0.55$ to $0.65$ across various nanowire morphologies (e.g., dispersed, percolated). Thus, the quantitative performance of our MTDN is well within the higher range of performance on similar problems.

Our quantitative results highlight MTDN’s advantage in segmenting smaller, more complex reactivity descriptors such as $A_2$. To further validate the comparative performance of MTDN and U-Net on the test set, we assessed their performance on an frame-by-frame basis over time, confirming that MTDN consistently slightly outperforms U-Net on $A_1$, and significantly outperforms U-Net on $A_2$. These results are illustrated in Figs.~\ref{fig:region6} and ~\ref{fig:region7} for Filament IDs 6 and 7, respectively. We note that both models initially achieve high F1 scores but experience a sharp decline after approximately 550 seconds. This decline occurs because $A_1$ and $A_2$ after 550 seconds are significantly smaller (indicating the filamentous carbon has been almost fully gasified) or almost empty, as illustrated in the Segmentation Visualization (right side of Figs.~\ref{fig:region6} and ~\ref{fig:region7}). Following convention, the F1 score for the missing class is set to 0 in this case, resulting in a much lower macro F1 score at the end of the gasification reaction.

\begin{table*}[!t]
\centering
\caption{Performance comparison for \textbf{\(\mathbf{A_1}\)} in the test dataset (Filament IDs 6 and 7). Higher performance is highlighted in bold.}
\resizebox{0.58\textwidth}{!}{
\begin{tabular}{|c|clllll|}
\hline
\multirow{3}{*}{Model} 
  & \multicolumn{6}{c|}{\textbf{\(\mathbf{A_1}\) Prediction}} \\ \cline{2-7} 
  & \multicolumn{3}{c|}{\textbf{F1 Score (Dice Score)}$\uparrow$} & \multicolumn{3}{c|}{\textbf{IoU (Intersection over Union)}$\uparrow$} \\ \cline{2-7} 
 &
  \multicolumn{1}{c|}{Filament ID 6} &
  \multicolumn{1}{c|}{Filament ID 7} &
  \multicolumn{1}{c|}{Test total} &
  \multicolumn{1}{c|}{Filament ID 6} &
  \multicolumn{1}{c|}{Filament ID 7} &
  \multicolumn{1}{c|}{Test total} \\ \hline
\multicolumn{1}{|l|}{\textbf{U-Net}} &
  \multicolumn{1}{l|}{0.90293} &
  \multicolumn{1}{l|}{\textbf{0.97206}} &
  \multicolumn{1}{l|}{0.94102} &
  \multicolumn{1}{l|}{0.83593} &
  \multicolumn{1}{l|}{\textbf{0.94673}} &
  0.89361 \\ \hline
\multicolumn{1}{|l|}{\textbf{MTDN}} &
  \multicolumn{1}{l|}{\textbf{0.91675}} &
  \multicolumn{1}{l|}{0.96746} &
  \multicolumn{1}{l|}{\textbf{0.9447}} &
  \multicolumn{1}{l|}{\textbf{0.85614}} &
  \multicolumn{1}{l|}{0.93844} &
  \textbf{0.89964} \\ \hline
\end{tabular}
}
\label{tab: A1 results}
\end{table*}

\begin{table*}[!t]
\centering
\caption{Performance comparison for \textbf{\(\mathbf{A_2}\)} in the test dataset (Filament IDs 6 and 7). Higher performance is highlighted in bold.}
\resizebox{0.58\textwidth}{!}{
\begin{tabular}{|c|clllll|}
\hline
\multirow{3}{*}{Model} &
  \multicolumn{6}{c|}{\textbf{\(\mathbf{A_2}\) Prediction}} \\ \cline{2-7} 
 &
  \multicolumn{3}{c|}{\textbf{F1 Score (Dice Score)}$\uparrow$} &
  \multicolumn{3}{c|}{\textbf{IoU (Intersection over Union)}$\uparrow$} \\ \cline{2-7} 
 &
  \multicolumn{1}{c|}{Filament ID 6} &
  \multicolumn{1}{c|}{Filament ID 7} &
  \multicolumn{1}{c|}{Test total} &
  \multicolumn{1}{c|}{Filament ID 6} &
  \multicolumn{1}{c|}{Filament ID 7} &
  \multicolumn{1}{c|}{Test total} \\ \hline
\multicolumn{1}{|l|}{\textbf{U-Net}} &
  \multicolumn{1}{l|}{0.76477} &
  \multicolumn{1}{l|}{0.7597} &
  \multicolumn{1}{l|}{0.76276} &
  \multicolumn{1}{l|}{0.67606} &
  \multicolumn{1}{l|}{0.66955} &
  0.67357 \\ \hline
\multicolumn{1}{|l|}{\textbf{MTDN}} &
  \multicolumn{1}{l|}{\textbf{0.85717}} &
  \multicolumn{1}{l|}{\textbf{0.8211}} &
  \multicolumn{1}{l|}{\textbf{0.8408}} &
  \multicolumn{1}{l|}{\textbf{0.77654}} &
  \multicolumn{1}{l|}{\textbf{0.73349}} &
  \textbf{0.75665} \\ \hline
\end{tabular}
}
\label{tab: A2 results}
\end{table*}

\subsection{Qualitative Results} \label{sec: qualitative Results}

\begin{figure*}[!h]
    \centering 
     \includegraphics[width=1\textwidth]
     {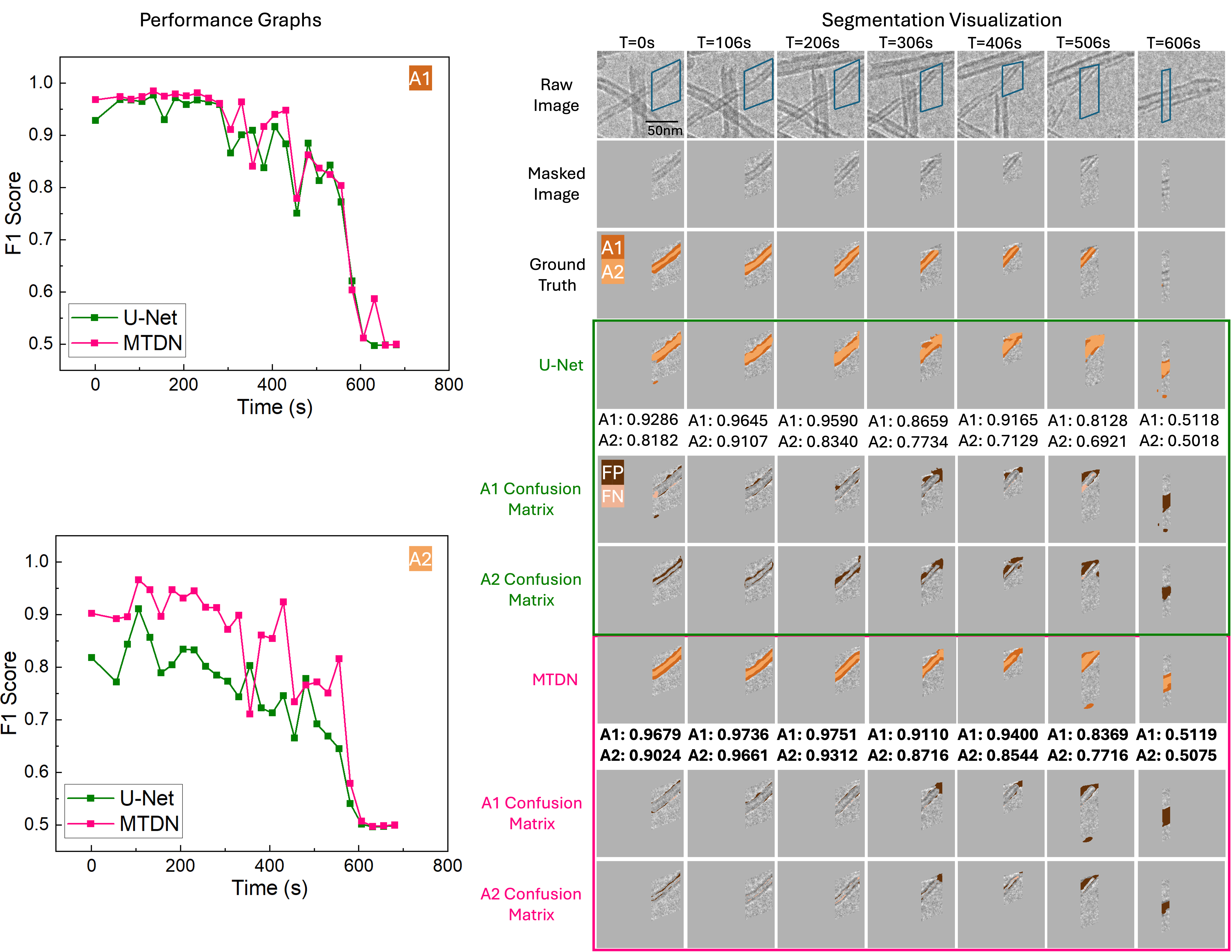} 
    \caption{
    F1 Scores over time (in seconds) for each model for Filament ID 6. In the Performance Graphs (left), the green line represents the performance of U-Net, while the pink line indicates the performance of MTDN. In the Segmentation Visualization (right), there are three rows for each model. The first of these rows contains the model predictions for A$_1$ and A$_2$, where A$_1$ includes both the dark and light orange regions, while A$_2$ corresponds to just the light orange region. The second and third rows correspond to the confusion matrices for the A$_1$ and A$_2$ predictions, respectively. In the confusion matrices, false positive (FP) regions are dark brown and false negative (FN) regions are light tan. The highlighted F1 scores demonstrate MTDN's consistently superior performance over time, particularly for $A_2$.} 
    \label{fig:region6}
\end{figure*}

\begin{figure*}[!h]
    \centering 
     \includegraphics[width=1\textwidth]{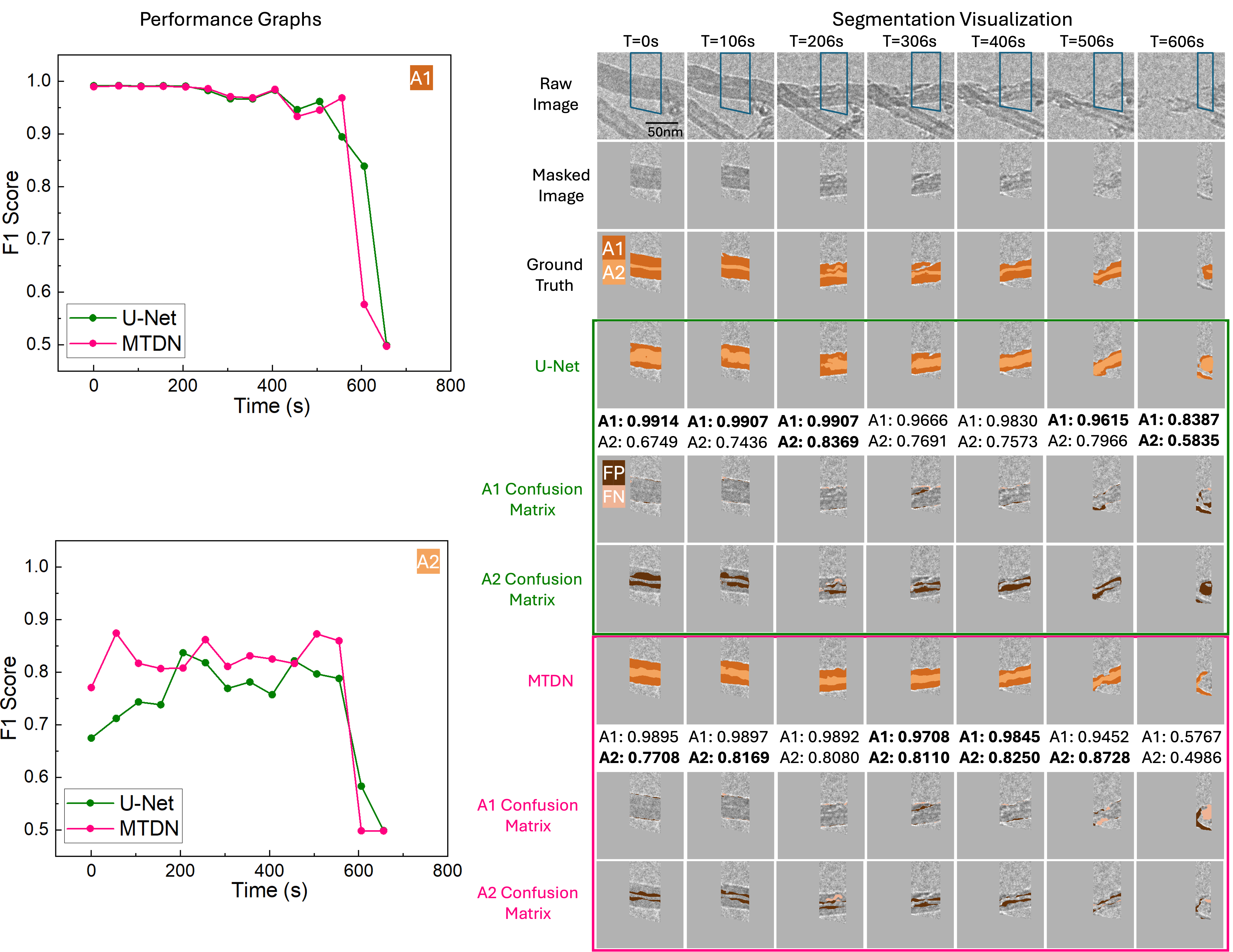}   
    \caption{F1 Scores over time (in seconds) for each model in Filament ID 7. In the Performance Graphs (left side), MTDN (pink) achieves a similar F1 score for A$_1$ when compared to U-Net (green). However, MTDN maintains consistently higher F1 scores for A$_2$. In the Segmentation Visualization (right side), we can see that MTDN shows better visual alignment with the ground truth.}
    \label{fig:region7}
\end{figure*}

Fig.s ~\ref{fig:region6} and ~\ref{fig:region7} also visualize the differences between U-Net and MTDN predictions of $A_1$ and $A_2$ over time for Filament IDs 6 and 7. The raw frames, masked frames, ground truth, segmentation predictions, and confusion matrices are shown for timesteps spanning early, middle and late times in the corresponding videos. Note that in the visualized predictions, the reactivity descriptor $A_2$ is highlighted in the lighter color, while the reactivity descriptor $A_1$ corresponds to both the dark and light colors combined. As the filamentous carbon shrinks, segmentation becomes more challenging for all models, but MTDN continues to perform consistently better. In particular, MTDN tends to produce more narrow, better predictions for $A_2$ compared to U-Net. This suggests that MTDN reduces false positive predictions, resulting in segmentation results that are more accurate and sharper, which is critically important for the area-to-volume conversion to quantify carbon volume changes for spatially-resolved operando ETEM characterization (Fig.~\ref{fig:experiment}).

For Filament ID 6 (Fig.~\ref{fig:region6}, we observe from time step 306 s to 406 s that U-Net too sensitive to differences in pixel intensity compared to MTDN. U-Net tends to predict all areas with contrast as belonging to \(A_1\), whereas MTDN is able to identify \(A_1\) more accurately. Additionally, from time step 306 s to 506 s, we encountered an overlap problem when an overlying carbon filament moved into proximity with the target filament, preventing us from fully excluding it by masking. U-Net incorrectly identifies all the carbon filaments as part of $A_1$. In contrast, our model accurately distinguishes between the two separate carbon filaments and to a good extent correctly predicts A$_1$ and A$_2$ belonging to the target filament. Finally, at time step 606 s, an additional lower carbon filament appears in the masked region, while the target filament has been completely gasified so that the ground truth label is empty. In this situation, both U-Net and MTDN fail to track the correct filaments (although visually we can see that MTDN better captures the \textit{new} filament). This failure reflects not a flaw of the model, but a necessary compromise to enable filament-specific gasification kinetic measurements that deconvolute the effects of size variation and distinct reaction pathways.

In Fig.~\ref{fig:region7}, we again see that MTDN consistently outperforms U-Net across the entire gasification process for Filament ID 7, both quantitatively and visually. Similarly to before, towards the end of gasification at time step 606 s, a new lower carbon filament appears while only a tiny portion of the target filament remains. U-Net tends to over-segment high-contrast areas, mistakenly identifying both carbon filaments. Meanwhile, MTDN predicts the emerging filament but omits the target one, resulting in MTDN having more false negatives than U-Net. In this instance, U-Net inadvertently achieves better quantitative results. 

Overall, the analysis of filaments 6 and 7 confirms that MTDN maintains strong and stable performance in complex segmentation scenarios, especially for the challenging $A_2$ class. Its ability to produce precise segmentations with fewer false positives and better temporal consistency highlights its robustness compared to U-Net.

\subsection{Ablation Study} \label{sec: ablation Study}
In this section, we study the relative importance of various components of our MTDN model: the performance gains from our multi-task formulation, the efficacy of pre-training the U-Net branches, and various fusion methods for transforming the change detection prediction to image segmentation results. 

\subsubsection{Multi-task Training}
We leverage the relationship between the $A_1$ and $A_2$ segmentation task to boost model performance using multi-task learning. In Table ~\ref{tab:multi-task}, we present comparative results between the results of the Multi-Task Detection Network (MTDN) with multi-task training against a MTDN model that only undergoes single-task training. The single-task training includes one MTDN model focused solely on predicting $A_1$ (left columns of table) and another MTDN model focused solely on predicting $A_2$ (right columns of table). The multi-task model clearly performs better than the single-task models.

\begin{table*}[!t]
\centering
\caption{Test performance of MTDN trained as a multi-task model versus MTDN trained as a single-task model on $A_1$ only or $A_2$ only.}
\resizebox{0.58\textwidth}{!}{
\begin{tabular}{|l|llllll|}
\hline
\multicolumn{1}{|c|}{\multirow{3}{*}{Model}} &
  \multicolumn{6}{c|}{\textbf{F1 Score (Dice Score)}$\uparrow$} \\ \cline{2-7} 
\multicolumn{1}{|c|}{} &
  \multicolumn{3}{c|}{\textbf{\(\mathbf{A_1}\) Prediction}} &
  \multicolumn{3}{c|}{\textbf{\(\mathbf{A_2}\) Prediction}} \\ \cline{2-7} 
\multicolumn{1}{|c|}{} &
  \multicolumn{1}{c|}{Filament ID 6} &
  \multicolumn{1}{c|}{Filament ID 7} &
  \multicolumn{1}{c|}{Test total} &
  \multicolumn{1}{c|}{Filament ID 6} &
  \multicolumn{1}{c|}{Filament ID 7} &
  \multicolumn{1}{c|}{Test total} \\ \hline
MTDN (Multi-task) &
  \multicolumn{1}{l|}{\textbf{0.91678}} &
  \multicolumn{1}{l|}{0.9694} &
  \multicolumn{1}{l|}{\textbf{0.94588}} &
  \multicolumn{1}{l|}{\textbf{0.8374}} &
  \multicolumn{1}{l|}{\textbf{0.82362}} &
  \textbf{0.83117} \\ \hline
MTDN (Single-task) &
  \multicolumn{1}{l|}{0.88336} &
  \multicolumn{1}{l|}{\textbf{0.97039}} &
  \multicolumn{1}{l|}{0.93014} &
  \multicolumn{1}{l|}{0.81126} &
  \multicolumn{1}{l|}{0.81154} &
  0.81152 \\ \hline
\end{tabular}
\label{tab:multi-task}
}
\end{table*}

\subsubsection{Pre-training of the U-Net Backbone}
We experimented with weight initialization for the Siamese branches using the pre-trained U-Net models based on \( A_1 \) (MTDN\_init$_1$) and \( A_2 \) (MTDN\_init$_2$). After this step, we continued end-to-end training of the full MTDN model, resulting in fine-tuning of these U-Net weights. Additionally, we conducted experiments without weight initialization.

\begin{figure*}[!t]
    \centering 
     \includegraphics[width=1\textwidth]{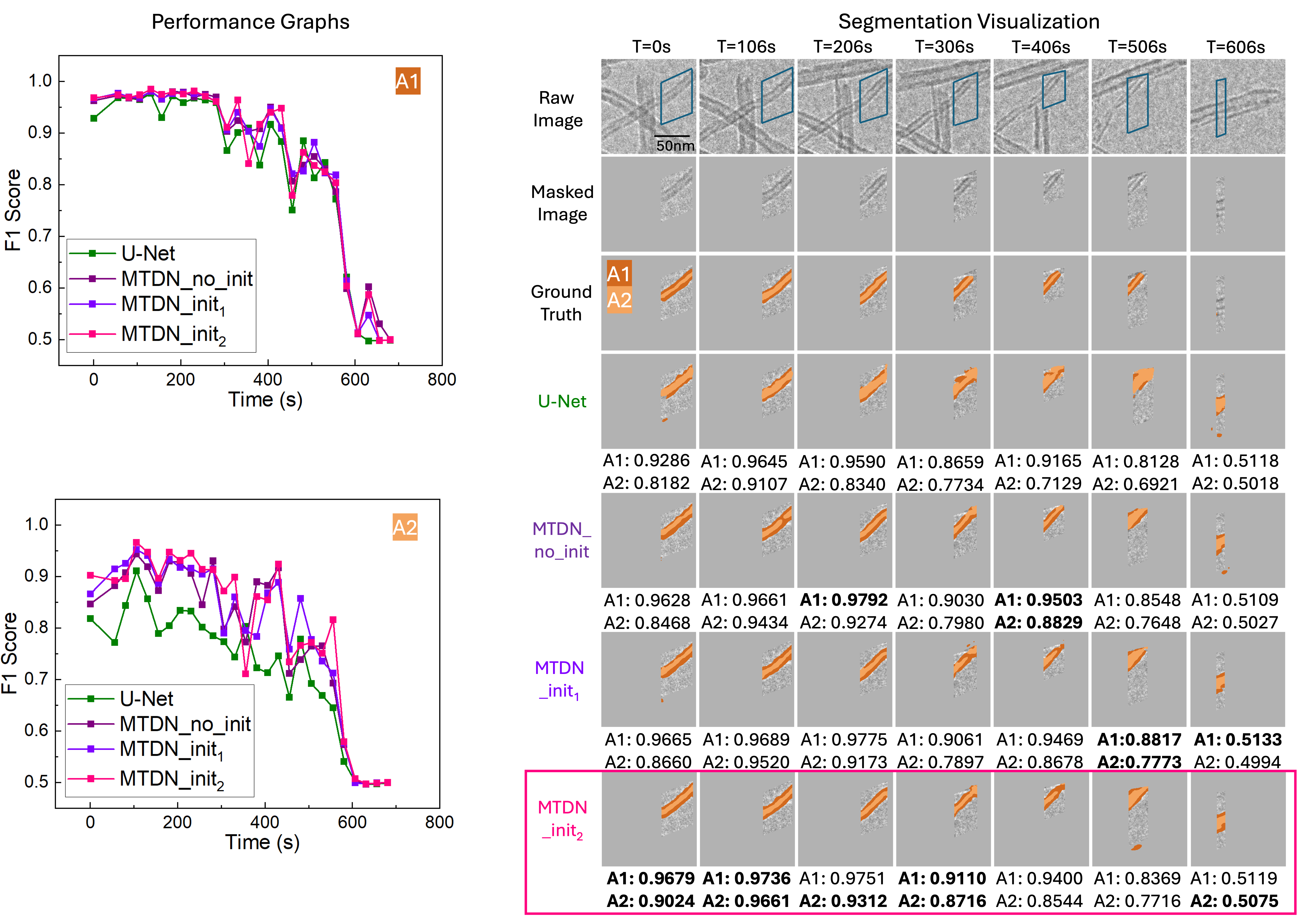} 
    \caption{F1 Scores and visualization of segmentation predictions over time for each model on Filament ID 6. The F1 scores for the best model at each timestep and each reactivity descriptor (A$_1$ or A$_2$) are highlighted in bold. The various models achieve similar performance over time. We select the MTDN\_init$_2$ model as the final MTDN model reported in the Results section, since higher performance on the challenging A$_2$ is more critical in this application.}
    \label{fig:201_test_all}
\end{figure*}

We present the numerical results in Table~\ref{tab:weights init} and the corresponding visualizations in Fig.~\ref{fig:201_test_all}. Our results indicate that pre-training the UNet to initialize our Siamese branches slightly enhances overall performance. Considering the overall F1 score, MTDN\_init$_2$ is the best model. This is the final MTDN model that is reported in the Quantitative and Qualitative Results sections. Tt is important to note that although we have employed a U-Net backbone for our Siamese branches, we can replace U-Net with any state-of-the-art encoder-decoder architecture of our choice in our MTDN model. We note that due to the small amount of labeled data, a lightweight backbone is preferred.

\begin{table*}[!t]
\centering 
\caption{Performance comparison of MTDN without weight initialization versus MTDN with U-Net, trained using $A_1$ weight initialization, and MTDN with U-Net, trained using $A_2$ weight initialization. }
\resizebox{0.58\textwidth}{!}{
\begin{tabular}{|l|lllllll|}
\hline
\multicolumn{1}{|c|}{\multirow{3}{*}{Model}} &
  \multicolumn{7}{c|}{\textbf{F1 Score (Dice Score)}$\uparrow$} \\ \cline{2-8} 
\multicolumn{1}{|c|}{} &
  \multicolumn{3}{c|}{\textbf{\(\mathbf{A_1}\) Prediction}} &
  \multicolumn{3}{c|}{\textbf{\(\mathbf{A_2}\) Prediction}} &
  Overall \\ \cline{2-8} 
\multicolumn{1}{|c|}{} &
  \multicolumn{1}{c|}{Filament ID 6} &
  \multicolumn{1}{c|}{Filament ID 7} &
  \multicolumn{1}{c|}{Test total} &
  \multicolumn{1}{c|}{Filament ID 6} &
  \multicolumn{1}{c|}{Filament ID 7} &
  \multicolumn{1}{c|}{Test total} &
  \multicolumn{1}{c|}{} \\ \hline
MTDN\_no\_init &
  \multicolumn{1}{l|}{0.91678} &
  \multicolumn{1}{l|}{\textbf{0.9694}} &
  \multicolumn{1}{l|}{0.94588} &
  \multicolumn{1}{l|}{0.838} &
  \multicolumn{1}{l|}{0.82295} &
  \multicolumn{1}{l|}{0.83117} &
  0.888525 \\ \hline
MTDN\_init$_1$ &
  \multicolumn{1}{l|}{\textbf{0.92036}} &
  \multicolumn{1}{l|}{0.96679} &
  \multicolumn{1}{l|}{\textbf{0.94601}} &
  \multicolumn{1}{l|}{0.8497} &
  \multicolumn{1}{l|}{\textbf{0.82484}} &
  \multicolumn{1}{l|}{0.83866} &
  0.892335 \\ \hline
MTDN\_init$_2$ &
  \multicolumn{1}{l|}{0.91675} &
  \multicolumn{1}{l|}{0.96746} &
  \multicolumn{1}{l|}{0.9447} &
  \multicolumn{1}{l|}{\textbf{0.85717}} &
  \multicolumn{1}{l|}{0.8211} &
  \multicolumn{1}{l|}{\textbf{0.8408}} &
  \textbf{0.89275} \\ \hline
\end{tabular}
\label{tab:weights init}
}
\end{table*}

\subsubsection{Change Detection to Segmentation Transformation (Prediction Fusion)}
\begin{figure*}[!t]
    \centering 
     \includegraphics[width=1\textwidth]{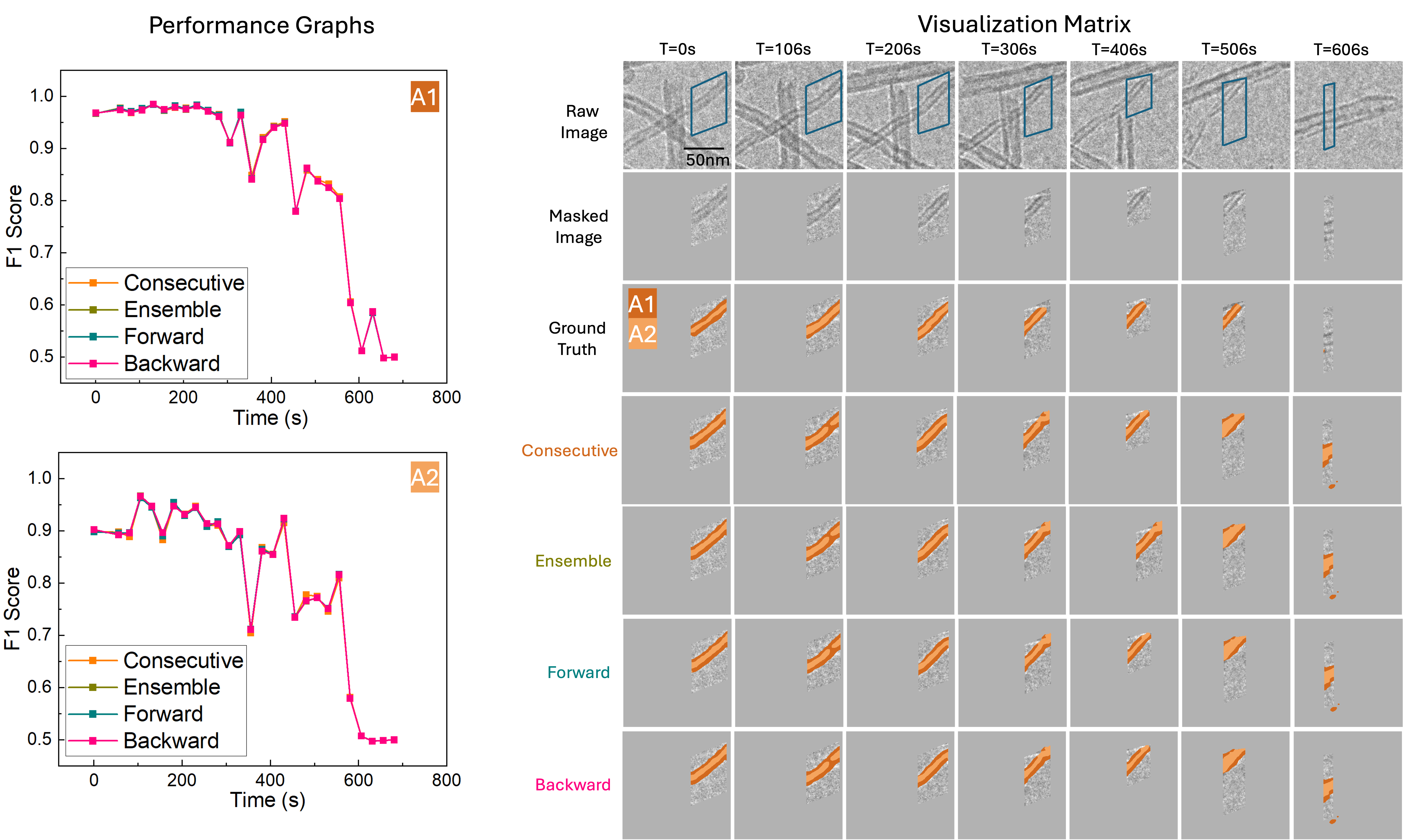}  
    \caption{Visual comparison of different prediction fusion methods on Filament ID 6. Visually, the results are quite similar.}
    \label{fig:region7_multi_method}
\end{figure*}

As discussed in Section~\ref{sec: area interest output}, a variety of fusion methods are possible to transform the change detection predictions to segmentation results. The quantitative performance and qualitative visualization of the different fusion methods are presented in Table~\ref{tab:multi methods} and Fig.~\ref{fig:region7_multi_method}. There is both high quantitative and visual similarity in the segmentation predictions across methods, suggesting that our MTDN framework is robust to the choice of prediction fusion method. For computational efficiency, we select MTDN\_init$_2$ using backward fusion as our final MTDN model for which the performance is reported in the Quantitative and Qualitative Results sections.

\begin{table*}[!t]
\centering
\caption{Performance comparison of different fusion methods for the change detection to image segmentation transformation for each model. The results are largely similar, showing that MTDN is largely robust to the choice of fusion method.}
\resizebox{0.80\textwidth}{!}{
\begin{tabular}{|l|l|lllllll|}
\hline
\multicolumn{1}{|c|}{\multirow{3}{*}{Model}} &
  \multicolumn{1}{c|}{\multirow{3}{*}{Fusion Method}} &
  \multicolumn{7}{c|}{\textbf{F1 Score (Dice Score)}$\uparrow$} \\ \cline{3-9} 
\multicolumn{1}{|c|}{} &
  \multicolumn{1}{c|}{} &
  \multicolumn{3}{c|}{\textbf{\(\mathbf{A_1}\) Prediction}} &
  \multicolumn{3}{c|}{\textbf{\(\mathbf{A_2}\) Prediction}} &
  Overall \\ \cline{3-9} 
\multicolumn{1}{|c|}{} &
  \multicolumn{1}{c|}{} &
  \multicolumn{1}{c|}{Filament ID 6} &
  \multicolumn{1}{c|}{Filament ID 7} &
  \multicolumn{1}{c|}{Test total} &
  \multicolumn{1}{c|}{Filament ID 6} &
  \multicolumn{1}{c|}{Filament ID 7} &
  \multicolumn{1}{c|}{Test total} &
   \\ \hline
MTDN\_no\_init &
  \textbf{backward} &
  \multicolumn{1}{l|}{0.91678} &
  \multicolumn{1}{l|}{\textbf{0.9694}} &
  \multicolumn{1}{l|}{0.94588} &
  \multicolumn{1}{l|}{0.838} &
  \multicolumn{1}{l|}{0.82295} &
  \multicolumn{1}{l|}{0.83117} &
  0.888525 \\ \hline
\textbf{} &
  \textbf{consecutive} &
  \multicolumn{1}{l|}{0.91801} &
  \multicolumn{1}{l|}{0.9691} &
  \multicolumn{1}{l|}{\textbf{0.94638}} &
  \multicolumn{1}{l|}{0.83083} &
  \multicolumn{1}{l|}{0.82597} &
  \multicolumn{1}{l|}{0.82873} &
  0.887555 \\ \hline
\textbf{} &
  \textbf{ensemble} &
  \multicolumn{1}{l|}{0.91699} &
  \multicolumn{1}{l|}{0.96894} &
  \multicolumn{1}{l|}{0.94566} &
  \multicolumn{1}{l|}{0.83812} &
  \multicolumn{1}{l|}{0.82491} &
  \multicolumn{1}{l|}{0.83216} &
  0.88891 \\ \hline
\textbf{} &
  \textbf{forward} &
  \multicolumn{1}{l|}{0.91798} &
  \multicolumn{1}{l|}{0.96865} &
  \multicolumn{1}{l|}{0.94604} &
  \multicolumn{1}{l|}{0.83591} &
  \multicolumn{1}{l|}{0.82599} &
  \multicolumn{1}{l|}{0.83146} &
  0.88875 \\ \hline
MTDN\_init$_1$ &
  \textbf{backward} &
  \multicolumn{1}{l|}{0.92036} &
  \multicolumn{1}{l|}{0.96679} &
  \multicolumn{1}{l|}{0.94601} &
  \multicolumn{1}{l|}{0.8497} &
  \multicolumn{1}{l|}{0.82484} &
  \multicolumn{1}{l|}{0.83866} &
  0.892335 \\ \hline
 &
  \textbf{consecutive} &
  \multicolumn{1}{l|}{0.92149} &
  \multicolumn{1}{l|}{0.96622} &
  \multicolumn{1}{l|}{0.94627} &
  \multicolumn{1}{l|}{0.84588} &
  \multicolumn{1}{l|}{0.82682} &
  \multicolumn{1}{l|}{0.83744} &
  0.891855 \\ \hline
 &
  \textbf{ensemble} &
  \multicolumn{1}{l|}{0.92042} &
  \multicolumn{1}{l|}{0.96598} &
  \multicolumn{1}{l|}{0.94554} &
  \multicolumn{1}{l|}{0.84982} &
  \multicolumn{1}{l|}{\textbf{0.82704}} &
  \multicolumn{1}{l|}{0.83976} &
  0.89265 \\ \hline
 &
  \textbf{forward} &
  \multicolumn{1}{l|}{\textbf{0.92163}} &
  \multicolumn{1}{l|}{0.96562} &
  \multicolumn{1}{l|}{0.94597} &
  \multicolumn{1}{l|}{0.84781} &
  \multicolumn{1}{l|}{0.82685} &
  \multicolumn{1}{l|}{0.83852} &
  0.892245 \\ \hline
MTDN\_init$_2$ &
  \textbf{backward} &
  \multicolumn{1}{l|}{0.91675} &
  \multicolumn{1}{l|}{0.96746} &
  \multicolumn{1}{l|}{0.9447} &
  \multicolumn{1}{l|}{0.85717} &
  \multicolumn{1}{l|}{0.8211} &
  \multicolumn{1}{l|}{0.8408} &
  \textbf{0.89275} \\ \hline
\textbf{} &
  \textbf{consecutive} &
  \multicolumn{1}{l|}{0.9184} &
  \multicolumn{1}{l|}{0.96661} &
  \multicolumn{1}{l|}{0.94512} &
  \multicolumn{1}{l|}{0.85479} &
  \multicolumn{1}{l|}{0.82291} &
  \multicolumn{1}{l|}{0.84034} &
  0.89273 \\ \hline
 &
  \textbf{ensemble} &
  \multicolumn{1}{l|}{0.91677} &
  \multicolumn{1}{l|}{0.96734} &
  \multicolumn{1}{l|}{0.94458} &
  \multicolumn{1}{l|}{\textbf{0.85722}} &
  \multicolumn{1}{l|}{0.82107} &
  \multicolumn{1}{l|}{\textbf{0.84085}} &
  0.892715 \\ \hline
 &
  \textbf{forward} &
  \multicolumn{1}{l|}{0.91738} &
  \multicolumn{1}{l|}{0.96708} &
  \multicolumn{1}{l|}{0.94481} &
  \multicolumn{1}{l|}{0.85569} &
  \multicolumn{1}{l|}{0.82105} &
  \multicolumn{1}{l|}{0.83996} &
  0.892385 \\ \hline
\end{tabular}
\label{tab:multi methods}
}
\end{table*}

\section*{Conclusions}

In this work, we introduce MTDN, a novel deep learning framework that reframes semantic segmentation as a change detection task, enabling spatially-resolved \textit{operando} ETEM characterization of filamentous carbon gasification, to accelerate the fundamental understand of catalyst regeneration. By leveraging a Siamese U-Net architecture that takes pairwise data as input, MTDN makes efficient use of limited training data, achieving high segmentation performance on complex reactivity descriptors. Critically, our prediction fusion methods convert change detection results to image segmentations quickly and efficiently, without suffering from accumulation of errors and without requiring additional manual labeling. Our model also benefits from a lightweight design, enabling flexible backbone replacement and efficient training. We further utilize a multi-task learning strategy to enhance the model’s ability to segment the two reactivity descriptors - the outer and inner regions of carbon filament - simultaneously. 

Extensive quantitative and qualitative analyses confirm that MTDN consistently outperforms traditional segmentation models such as U-Net, particularly in generalization and robustness across temporal variations and structural complexities. Ablation studies validate the impact of multi-task training, weight initialization with fine-tuning, and segmentation prediction fusion strategies, underscoring the effectiveness of our architectural and methodological choices.

Overall, MTDN accelerates the transformation of conventional in-situ (E)TEM imaging into spatially-resolved \textit{operando} (E)TEM characterization, by offering an automated approach to track the spatiotemporal evolution of (nano)materials with unprecedented speed, precision, and statistical rigor. This advance opens tremendous opportunities for mechanistic studies of solid-state reactions, where feature-specific reaction kinetics resolved at the nanometer scales can be directly correlated with its microstructural evolution. Looking ahead, this framework establishes a  strong foundation for future research in deep learning-driven microscopy, particularly in domains where labeled data is scarce and small objects are inherently present.

\section*{Data Availability}
Data and source code for this article, as well as scripts to reproduce experiments, will be made available in a public GitHub repository associated with this paper upon publication.

\section*{Conflicts of Interest}
There are no conflicts to declare.

\section*{Author contributions}
\credit{Yushuo Niu}{Data curation, Formal analysis, Investigation, Methodology, Software, Validation, Visualization, Writing -- original draft}
\credit{Tianyu Li}{Data curation, Formal analysis, Investigation, Methodology, Software, Validation, Visualization, Writing -- original draft}
\credit{Yuanyuan Zhu}{Conceptualization, Methodology, Funding acquisition, Project administration, Resources, Writing -- review \& editing}
\credit{Qian Yang}{Conceptualization, Methodology, Funding acquisition, Project administration, Resources, Supervision, Writing -- review \& editing}

\section*{Acknowledgements}
Y. N. and Q. Y. are supported by funding from the National Science Foundation under Grant No. DMR-2102406. This material is based upon work supported by the U.S. Department
of Energy’s Office of Energy Efficiency and Renewable Energy (EERE) under the
Hydrogen and Fuel Cell Technologies Office (HFTO) Award Number DE-EE0011303. The views expressed herein do not necessarily represent the views
of the U.S. Department of Energy or the United States Government. T. L. and Y. Z. are supported by the Catalysis Program, in the Division of Chemical, Bioengineering, Environmental, and Transport Systems of the United States National Science Foundation under award NSF CBET-2238213.


\balance

\renewcommand\refname{References}

\bibliography{bibnotes} 
\bibliographystyle{rsc} 
\end{document}